# Neurological Nature of Vision and Thought and Mechanisms of Perception Experiences

**Jahan N Schad***

*Retired LBNL (UCB) Scientist, USA*

**\*Corresponding author:** Jahan N Schad, PhD., Retired LBNL (UCB) Scientist, 376 Tharp Drive, Moraga, Ca 94556, California, USA, Tel: 925-376-4126; Email: Jaschadn@gmail.com

**Abstract**

Understandings of the phenomena of *vision* and *thought* require clarification of the general mechanism of perception, -the experience prompted by the (brain) efferent signals -as well as the clarification of the natures of the related afferent signals, which drive the mechanism. So far, philosophical inquiries and scientific investigations have not been able to address clearly the mysteries surrounding them.

The present work is an attempt to unravel the essences of these phenomena based on the presumption of computational functioning of the brain, a concept supported by scientific consensus. Within this context, the nature of the thought is clarified, and the basis of the experience of perception is established. And by drawing from the successes of the tactile vision substitution system (TVSS) [1], -- which renders a measure of vision perception in vision handicapped, early or congenital blinds -- the true nature of the vision, as cutaneous sensation, is also divulged.

The mechanism of perception --what renders it and where it occurs --involves sensing of stimuli, and, or, the autonomous engagement of brain inherent neuronal complexity resolution patterns; the implicit embedded computational instruction (codes). Upon commencement of such triggers, --of which one may not be necessarily aware --brain computations, which also involve engaging body's biophysiological feedback system, are performed; and the results are outputted as motor (efferent) signals that render perception. However, embedded in the process of the development of the experience of perception, is the deployment of a perception medium; an interface. Given the nature of the efferent signals, there must be a (known) biomechanical system interface, --other than the irrelevant body muscle and skeletal systems --which performs the needed function: Considering the fact that the vocal system performs such task for the verbalization of brain's synthesis of language expressions, the possibility of its further role in the experiences of thought and vision, in the form of mostly quiet (inaudible) recital of related signals, is suggested.

## Hypothesis

Understanding of the phenomenon of *Vision,* which is to know "how and where we see what we see," beyond the knowledge of the biophysiological and optical aspects of the eyes, and the brain modalities where the trigger signals are processed, has remained a mystery. Neurosciences' knowledge of the central nervous system, and the brain neurocomputational concepts, suggest that brain neuronal code (computational patterns) processing of eye-extracted afferent data (somehow) renders vision perception. However, this still falls short of a complete and convincing addressing of the above question; leaving the vision concept vague, as it has always been. Further, and very detailed understanding of the anatomy and physiology of the vision and the related processes [2], are not likely to provide the answer. We resolve this ambiguity by bringing to light the nature of the vision afferent) signals, their processing in the brain, and where the (brain) efferent vision signals are experienced.

Same difficulty for the *though*, as to "what it is and where it happens," has also always held true; beyond platitude. The philosophical addressing of mental processes and scientific understandings of the brain and its functions has not helped to resolve the puzzle either. A step toward the resolution of the thought ambiguity can be found in the biologic theory of linguistics, which according to Chomsky [3] entails the presence of neuronal language construct in the brain, and the proposition of two related interfaces: The first is "the *thought system* which provides *a place* for the interpreted Internal (I) language mechanism synthesis of the structured expression in the brain;" and the second is the "*vocal system*, activated by the motorsensory neurons," which renders language vocalizations; whether it is referential as in the calls of animals, or verbalized as in humans. However, despite this enlightening concept, the dilemma about the overall nature of the thought and its system still remains.

To address these mysteries, we focus on:

a) The nature of the brain information processing schemes and the triggers which drives them; and

b) On the nature of the brain outputs and the need for biological *interfaces*.

The presumption of computational functioning of the brain is based on the vast body of computational sciences and neurosciences findings, in tandem with experimental works in the area of information processing of the neurons [4]; and the successes of the brain inspired scientific neural networks in developing some measure of human-like intelligence. Extending the general workings principal of the artificial neural nets to the brain neuronal computations [5], is a very plausible assumption: In the scientific neural networks the resolution of complex problems calls for increasing units (layers) of calculation nodes, and verifiable [6] brain neuronal plasticity allows for engagement of various available neuronal modalities (possibly in hierarchical manner). Obviously brain neuronal net with the estimated availability of many trillions of biological microprocessors is an unfathomable complex, intelligent parallel process computation engine that is evolutionary perfected and configured for sustenance of life. Implications are its lightening speed, complexity resolution potential by virtue of its constructs, and the engrained learned schemes (as neural patterns), for handling all relevant natural phenomena; including vision and thought that life entails.

The process of natural phenomena resolution in the brain begins with the receipt of the sensed stimuli, *and/or* with the autonomous deployment of brain inherent computational neuronal patterns (implicit complexity resolution codes); which is followed by the onset of necessary computations while engaging body's biophysiological feedback system. Clearly, these operations, due to the ever presence of triggers, are perpetual; and the streaming outputs, as motor (efferent) signals, render continuous perceptions of various phenomena. Among them are the experience of *thought*, triggered by internal or external environmental elements; and the experience of *vision*, mostly by the external environmental triggers during waking hours; of which one may or may not be aware. However, embedded in the process of the experience of perception, is the availability of a medium for it; an interface. Given the nature of the efferent signals, there must be a known *biomechanical system interface*, other than body muscle and skeletal systems, which performs the needed function.

The vocal system, mentioned earlier, is a proven candidate: This interface is responsible for the vocalization of the language, and occasionally of thought; the latter true for almost all. And this experience of switching from thought to talk, metaphorically a gearshift, discloses perhaps disruptively, *the immense possibility of presence of dual mode to the vocal system, which allows for expressions of audible and inaudible thoughts.*

As the to natures of vision efferent signals, we made a seemingly important and unexpected discovery by critically examining the experimental results of the tactile vision substitution system (TVSS), demonstrated in the initial work of Bach-y-Rita et al. [1], Published in Nature: The work had established the development of vision-like perception in blind subjects, when fitted with the system which includes a pulsating patch on the skin or tongue. Taking note of the fact that such subjects never experience vision, neither in waking hours nor in dreams [7], the experience of such perceptions seemed inexplicable: On the face of it, the patch should only create cutaneous perceptions. True that from the patch location, massive and simultaneous amount of data pulses are sent to the brain, however, this should only lead to the development of some *matter (object)* perception, likely that of the patch itself. However, the repeated experimentally verified phenomenon can be explained in the context of brain neuronal net computational procedure which can engage any available neural circuitry (brain modality), including those of vision, for the processing of large amount of afferent data, such as those of cutaneous nature sensed from the TVSS patches. Given the visual perception experiences of the blind experimental subjects, rendered by the brain post processed efferent signals, the inference would be that the difference between the natures of the vision and cutaneous afferent signals must not be in kind but only in intensity, which dictates the vision details. And this claim is evinced by the hazy visual-like perceptions of blind subjects due to still insufficiency of the tactile afferent data. Based on such observations the disruptive discovery is that: *the nature of vision perception is same as cutaneous perception, and that there should not be any differences in the natures of their afferent signals.*

As in the case of thought, perception experiences require an interface, a display venue for the related brain (output) efferent signals-- generated post computational processing of the environmental stimuli from the vision and tactile sensing: A known human biological systems interface must be serving this function: Considering the fact that the vocal system performs such a task for the verbalization of brain's synthesis of language expressions efferent signals, the possibility of its versatility for expression of the thought, the tactile and the vision perception experiences in the form of quiet (inaudible) -- often unaware -- recital of related information, is suggested.

## Validation

A measure of validation can be found in the very comprehensive and detailed experimental work on mirror neuron activities performed by Keysers et al. [8,9]: They examined a phenomenon called "tactile sympathy." In these studies, the areas of motor neurons activated in a group of subjects watching a movie of a second group being very lightly touched on the skin were significantly similar to those who were actually being touched. Also the combined fMRI and TMS evidence of such sympathy, which is shown in the work of Alaert et al. [10], provides additional credence to our hypothesis. We believe these results provide strong experimental support for the concept of the tactile nature of vision presented in this theory.

For further validation of theory, we suggest experiments in which specific vocal motor neuron activities are monitored in different groups of subjects, some normally sighted and others with congenital blindness. Subjects would be monitored during speech, thinking, writing and visual (or, in the case of blind subjects, vibro-tactile) engagements. Vocal vibrations during conscious thinking would also be recorded. The results of such experimentation would definitively either prove or refute the theories put forward in this work.

## Conclusion

Visual and cutaneous stimuli sensations are (computationally) processed in the brain similarly; which are evinced by the development of vision perception in blind subjects, congenital or otherwise, fitted with tactile visual substitution (TVSS) systems. It is the scarcity of normal tactile sense data in blinds, which limits their proper perceptions of the environment. In case of normal eyesight, retinal neurons figuratively extend themselves by virtue of receiving rays of Photons which are environmentally modulated for the physical reality of the object from which they are reflected. Putting it simply, in the experience of vision we are being touched by the external world, while in cutaneous experience we are physically touching them.

Brain's computational operations are also constantly triggered by beings, exposure to other life phenomena, which are resolved, and streamed as efferent signals for perception. The experiences of perceptions in response to senses stimuli must be realized at a venue, a biomechanical interface which expresses (displays) the corresponding post process brain efferent signals: Vocal system, which serves language verbalization, is seemingly the only such device which could offer this possibility. And this leads to the presumption that vision, thought and tactile perceptions are but mostly inaudible utterances at the vocal machinery.

Perhaps this discovery would be disturbing to the poetic thought; on the face of it; however, knowing that *seeing* has more to it than sweep of glance, the "Thought" would be more incensed of its romantic implications!

## Acknowledgement

Thanks are due to Professor Noam Chomsky for his pioneering biolinguistic work and for the lecture that has served as an overall inspiration for me. Further thanks go to Dr. Roya Noorishad, who instigated my interest in neurosciences.

## References

1. Bach-y-Rita P, Collins CC, Saunders FA, White B, Scadden L (1969) Vision substitution by tactile image projection. Nature 221(5184): 963-964.

2. Schwarz SH (1999) Visual Perception. McGraw Hill, New York, USA.

3. Chomsky N (2007) On Language. The New Press, New York, USA.

4. Kandel ER, Schwartz JH, Jessell TM (2010) Principles of Neurosciences. McGraw-Hill, New York, USA.

5. Schad NJ (2016) Brain Neurological Constructs: The Neuronal Computational Schemes for Resolution of Life's complexities. J Neurol Neurophysiol 7(1): 356.

6. Edelman GM (1987) Neural Darwinism. The Theory of Neuronal Group Selection. Basic Books, New York, USA.

7. Edison, Tommy (2013) Intangible Concepts to a Blind Person. Youtube Recording.

8. Keysers C, Wicker B, Gazzola V, Anton JL, Fogassi L, et al. (2004) A touching sight SII/PV activation during the observation and experience of touch. Neuron 42(2): 335-346.

9. Keysers C, Gazzola V (2009) Expanding the mirror: Vicarious activity for actions, emotions and sensations. Curr Opin Neurobiol 19(16): 666-671.

10. Alaerts K, Swinnen SP, Wenderoth N (2009) Is human primary motor cortex activated by muscular or direction-dependent features of observed movements? Cortex 45(10): 1145-1155.